\documentclass[a4paper,12pt]{article}
\usepackage{graphicx}
\usepackage{setspace}
\usepackage{amsmath}
\usepackage{amssymb}
\usepackage{vmargin}
\setpapersize{A4}
\setmarginsrb{20mm}{15mm}{20mm}{15mm}{14.5pt}{8mm}{0pt}{11mm}

\begin{document}

\title{Polypropylene Embedded Metal-Mesh Broadband Achromatic Half Wave Plate for Millimeter Wavelengths}

\author{Zhang~J.$^1$\thanks{E-mail:
jin.zhang@astro.cf.ac.uk}, Ade~P.~A.~R.$^1$, Mauskopf~P.$^1$,
Savini~G.$^2$, Moncelsi~L.$^1$,
Whitehouse~N.$^1$\vspace{0.5cm}\\$^{1}$ School of Physics and
Astronomy, Cardiff University, The Parade,\\CF24 3AA Cardiff,
Wales, UK\\$^{2}$ Optical Science Laboratory, Physics and
Astronomy Department,\\University College London, Gower Street,
London WC1E 6BT, UK}

\date{}
\maketitle

\begin{abstract}

We describe a novel multi-layered metal mesh achromatic half wave
plate for use in astronomical polarimetric instruments. The half
wave plate is designed to operate across the frequency range from
125--250\,GHz. The wave plate is manufactured from 12-layers of
thin film metallic inductive and capacitive grids patterned onto
polypropylene sheets, which are then bonded together using a hot
pressing technique. Transmission line modelling and 3-D
electromagnetic simulations are used to optimize the parameters of
the metal-mesh patterns and to evaluate their optical properties.
A prototype half wave plate has been fabricated and its
performance characterized in a polarizing Fourier transform
spectrometer. The device performance is consistent with the
modelling although the measured differential phase shift for two
orthogonal polarizations is lower than expected. This difference
is likely to result from imperfect patterning of individual layers
and misalignment of the grids during manufacture.

\end{abstract}

\section{Introduction}

A wave plate is an optical component which imparts a differential
phase shift to light passing through it in different linear
polarization directions. This changes the polarization state of a
transmitted wave according to the magnitude of this angular phase
difference. A half wave plate (HWP) imparts a 180$^{\circ}$
differential phase shift between light with orthogonal electric
field components and can be used to rotate the polarization vector
by an angle, $2\theta$ by rotating the HWP by an angle $\theta$
with respect to the incident radiation. HWPs are used as
polarization modulators in astronomical instruments at millimeter
wavelengths ($\lambda$ = 1--3\,mm), in particular for measurements
of the residual polarization in Cosmic Microwave Background (CMB)
anisotropies \cite{Savini2006,Pisano2006}. These measurements
require large spectral bandwidths, cryogenic detectors, and
polarimeters with low instrumental polarization and low
instrumental emission to ensure photon noise limited detection of
the weak polarized CMB anisotropies. This has motivated the
development of large area achromatic HWPs
\cite{Savini2006,Pisano2006}.

Conventional HWP technologies use birefringent crystalline
materials such as Quartz or Sapphire to provide the differential
phase delay of orthogonal polarization components. While single
crystal plates demonstrate good performance in a narrow frequency
band, a broad-band design was reported initially for use at
optical wavelengths \cite{Pancharatnam1955,Rosenberg1981}. This
design incorporates 3 or 5 crystal plates of the same thickness,
but differentially rotated with respect to the incoming polarized
beam to provide a wider range of wavelengths over which the
retardation between orthogonal beams is constant. The first
achromatic multiple plate HWPs fabricated for a millimeter wave
CMB experiment \cite{Ade2008} used a sapphire 5-plate design to
cover the frequency range from 85--185\,GHz
\cite{Savini2006,Pisano2006}. This design has been shown to
perform well but is limited in aperture size (290\,mm for sapphire
\cite{Matsumura2009}) by the availability of materials.
Furthermore, the required HWP thickness ($\approx$17\,mm) leads to
a small absorption loss (3\,\%) which becomes a significant source
of radiative power and therefore photon noise compared to the
2.73\,K CMB, if the HWP is used at 300\,K. Lowering the
temperature to $\approx$100\,K eliminates this emission but
significantly complicates the instrument design and cost.

To overcome both the limited diameter and the absorption losses in
crystalline wave plates, an alternative method of making HWPs
based on metal-mesh technology has been proposed
\cite{Shatrow1995,Pisano2008}. This design uses capacitive and
inductive metal-mesh geometries which each generate a frequency
dependent phase shift with opposite sign. A first device
fabricated from 6 capacitive grids and 6 inductive separated by
air/vacuum gaps using accurately etched annular spacers
demonstrated a bandwidth of 40\,\% (60\,GHz) with a center
frequency of 150\,GHz \cite{Pisano2008}. Here we describe a
procedure for optimizing the design of this type of HWP and
present measurements of a device fabricated with a new process
using dielectric spacers between the metallized sheets fused
together with a hot pressing technique to make a solid
self-supporting disc \cite{Zhang2009}. This leads to a robust and
easily formed component with small absorption losses and a
diameter limited by the available size of the patterned meshes
which can be manufactured with diameters $\geq$300\,mm
\cite{Ade2006}.

In Section 2 of this paper we describe the theory of operation of
the HWP and the optimization procedure. We also present the design
parameters and expected performance of a HWP optimized for the
125--250\,GHz region. In section 3 we describe the device
manufacture and in section 4 we present optical performance
measurements from its characterization as a HWP using a polarizing
FTS.

\section{Theory and Modelling}

Wave plates based on multiple metal-mesh layers have been built by
Lerner \cite{Lerner1965}, Shatrow \cite{Shatrow1995} and Pisano
\cite{Pisano2008}. All of these designs incorporate a combination
of two basic types of metal patterns shown in Figure 1. The first
pattern consists of continuous parallel metal lines aligned along
one polarization axis. A single layer of this pattern acts like an
ideal inductive load for incident radiation with electric field
polarized parallel to the direction of the lines. The value of the
effective inductance depends on the width and spacing of the
lines. The second pattern consists of parallel metal lines with
small periodically spaced gaps. These lines are aligned along the
other polarization axis perpendicular to the first pattern. This
pattern acts like a combination of an inductive and a capacitive
load in series for incident radiation with the electric field
parallel to the direction of these lines. The values of the
inductance and capacitance of this pattern also depend on the
detailed geometry.

\begin{figure}[h!]\label{Fig-1}
\centering
\includegraphics[width=12cm,height=8cm,angle=0]{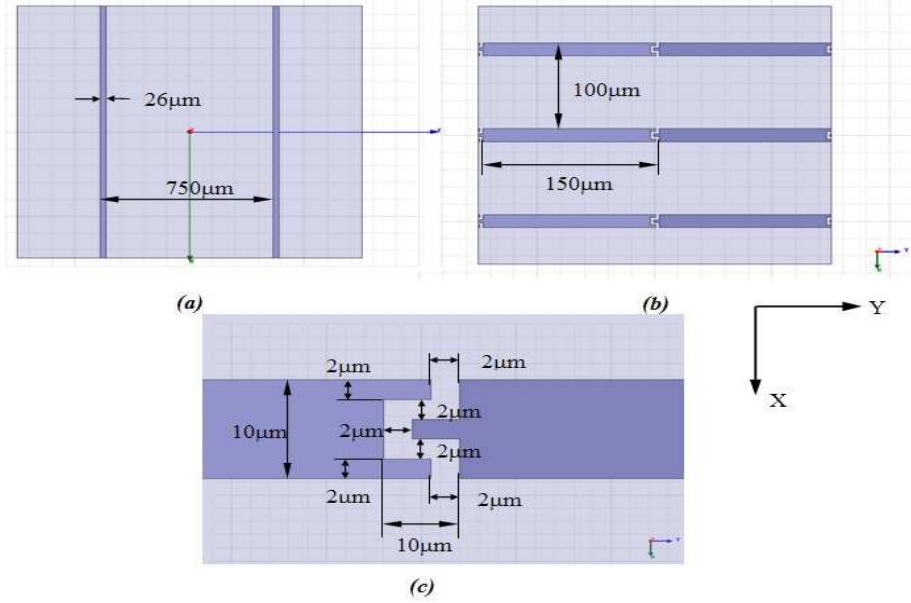}
\caption{Top view of single layer metal mesh pattern. (a)
Inductive grid pattern. (b) Capacitive grid pattern. (c) Metallic
planar interdigital capacitor in capacitive grid pattern.}
\end{figure}

We perform simulations and optimization of this design using two
commercially available modelling packages: Advanced Design System
(ADS \cite{ADS}) and High Frequency Structure Simulator (HFSS
\cite{Ansoft}). We use ADS for transmission line modelling and
optimization of a schematic model of the HWP using lumped
inductances and capacitances to represent the individual metal
layers. Based on the optimized values of the inductances and
capacitances determined from ADS, we use HFSS to relate the
geometrical parameters of an individual mesh to its lumped
impedance by breaking the physical mesh into cells and solving
Maxwell's equations on a cell by cell basis and thus obtaining the
scattering matrices for radiation propagation through the mesh.

We construct two independent transmission line models in ADS to
compute the scattering parameters for the two orthogonal
polarizations as shown in Figure 2. The first model considers the
transmission of the six aligned inductive grids with appropriate
polypropylene dielectric spacers between the grids. The second
model considers transmission of the six capacitive grids with
polypropylene spacers. For each transmission line model we add an
additional section of polypropylene designed to make the total
physical thickness of the two transmission line models equal to
the sum of the thicknesses of the two sets of six grids and
spacers. The scattering parameters from both transmission lines
are combined to determine the phase difference between the two
polarizations. This is a good approximation to the expected
performance of the HWP assuming: i) the grids designed to interact
with one polarization do not affect the other polarization and ii)
the grids do not generate cross-polarization - i.e the scattering
parameters between one transmission line and the other
transmission line are assumed to be equal to zero.

\begin{figure}[h!]\label{Fig-2}
\centering
\includegraphics[width=12cm,height=8cm,angle=0]{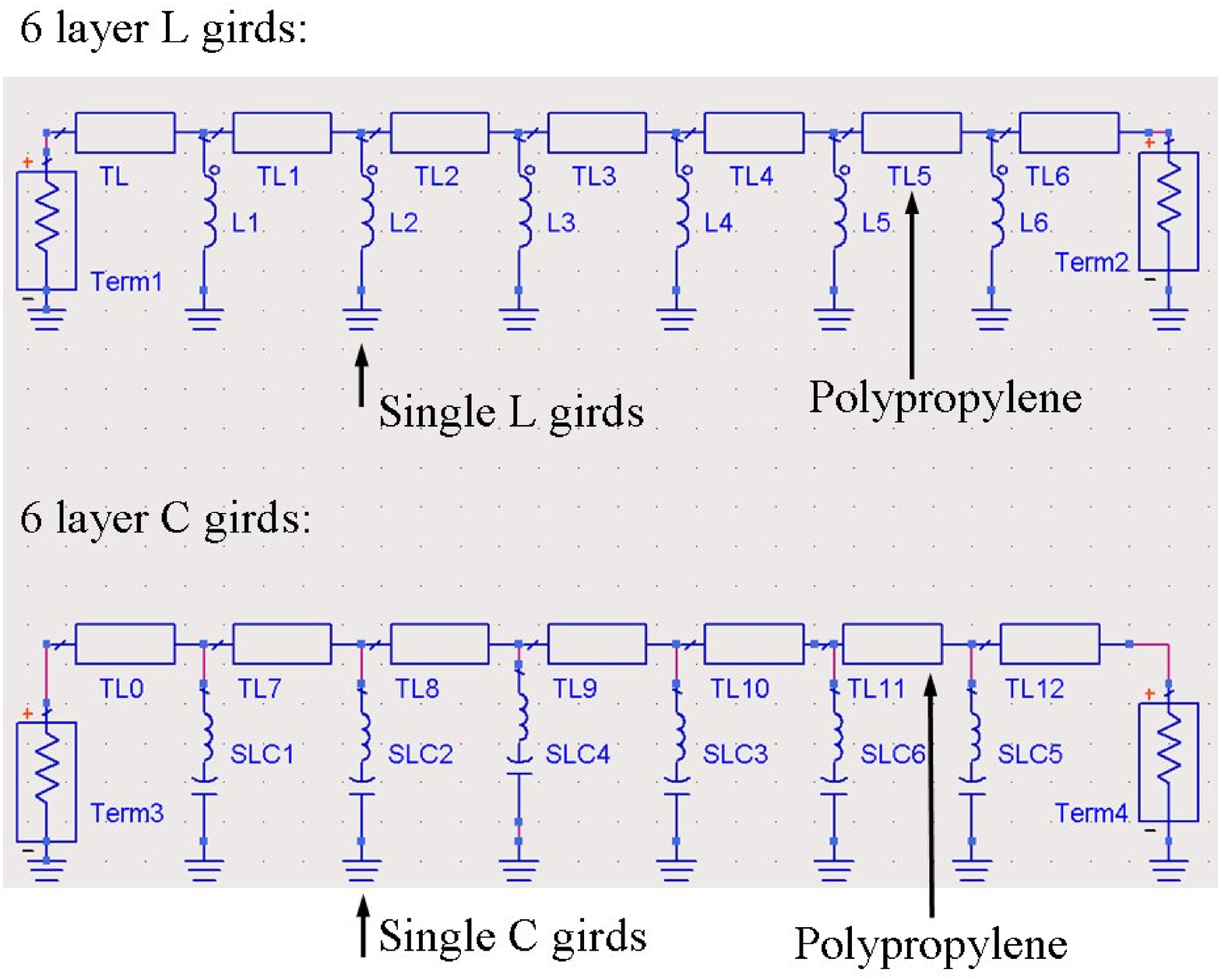}
\caption{Transmission line models for six capacitive grids and six
inductive grids embedded in polypropylene. An additional section
of polypropylene is added when modeling to make the total physical
thickness of the two transmission line models equal to the sum of
the thicknesses of the two sets of six grids and spacers.}
\end{figure}

The performance of the HWP schematic is optimized in ADS by
varying the values of the inductors and capacitors to achieve a
flat phase shift near $180^{\circ}$ and to maximize the
transmission over the required frequency range from 125 to
250\,GHz. This process combines the optimization of the relative
phase delay between orthogonal polarizations and a high broad band
transmission based on multiple reflections between the metal
layers. The optimized transmission for six inductive grids and six
capacitive grids are shown in Figures 3 and are seen to both be in
excess of 90\,\% throughout the band. For the prototype discussed
here the optimized HWP design gives an overall phase shift between
$170^{\circ}$ - $180^{\circ}$ as shown in Figure 4. Generally, we
find that achieving a broad bandwidth requires a high effective
capacitance and inductance in the ADS model.

\begin{figure}[h!]\label{Fig-3}
\centering
\includegraphics[width=12cm,height=8cm,angle=0]{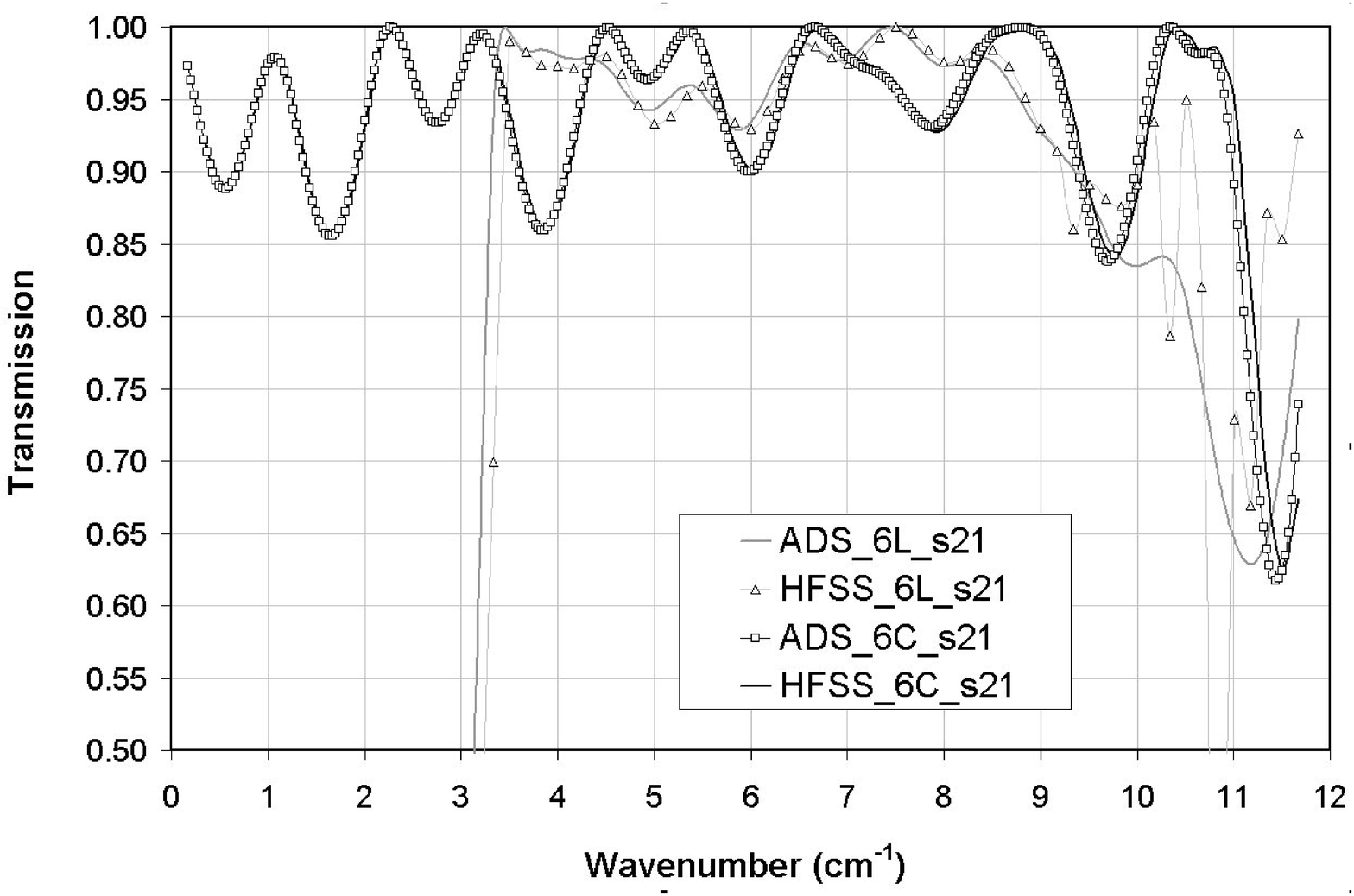}
\caption{HFSS and ADS simulated power transmission through HWP
when the electric field is parallel to the inductive lines
(inductive grids) and orthogonal to it (capacitive grids).}
\end{figure}

\begin{figure}[h!]\label{Fig-4}
\centering
\includegraphics[width=12cm,height=8cm,angle=0]{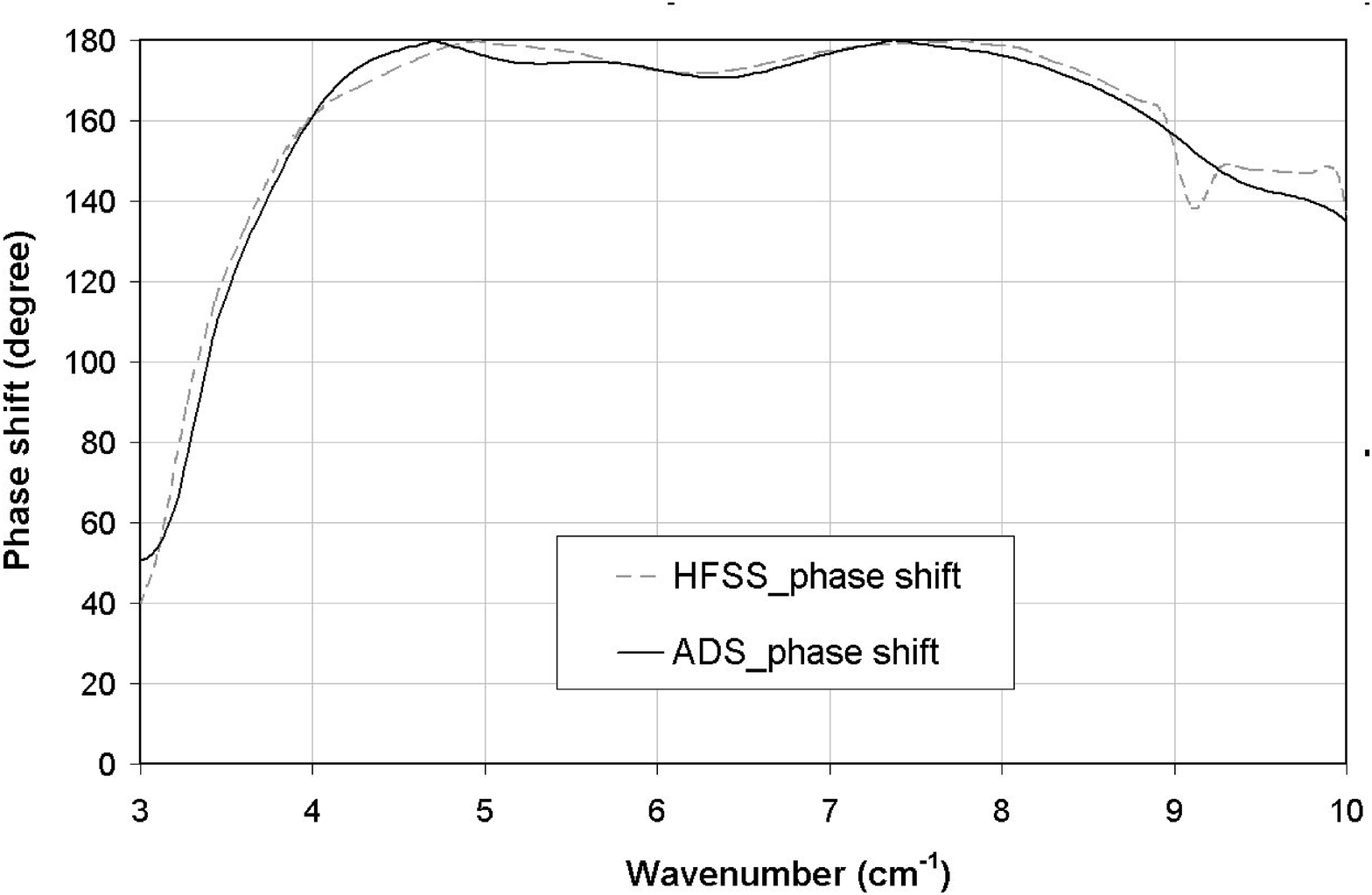}
\caption{HFSS and ADS simulated phase shift through HWP.}
\end{figure}

Based on the optimized lumped impedances determined from the
schematic model in ADS, we determined the corresponding grid
geometries by modelling the scattering parameters of individual
metal patterns in HFSS. The geometries of the inductive and
capacitive grids are shown in Figure 1. To achieve the high
effective lumped capacitance in the optimized ADS design we used a
periodic array of metallic planar interdigital capacitor coupled
lines as shown in Figure 1. The optimized design also required
high inductance which we realised by using very thin planar
parallel metallic lines. The frequency-dependent transmission and
phase shift for the individual inductive and capacitive grid
patterns as determined from ADS and HFSS models are shown in
Figures 3 and 4. This indicates that both the parallel inductive
wires and the interdigital capacitors are able to be accurately
modelled as lumped circuit elements over a wide frequency range.
This is not true for distributed resonators used as capacitive
elements in the previous air gap HWP design. In addition, the use
of non-resonant interdigital capacitors allows the grid dimensions
to be scaled independent of the wavelength. Because the spacings
between the layers are still designed to be resonant, the overall
HWP design is a hybrid between a so-called resonant filter and a
metamaterial. The metal strips in the model are simulated as
perfect conductors with zero thickness. In practice, the metal
layers are 0.4--1\,$\mu$m thick copper. We also perform
simulations using an infinitely thin conductor with a sheet
resistance corresponding to the measured value
($\approx$0.05\,$\Omega/\square$).

\begin{figure}[h!]\label{Fig-5}
\centering
\includegraphics[width=12cm,height=8cm,angle=0]{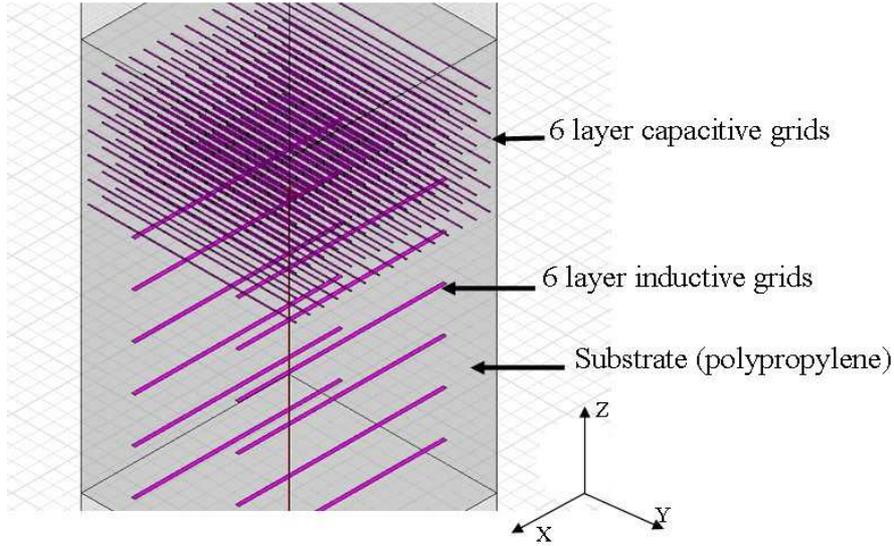}
\caption{The basic structure of the HWP model. There are twelve
grids embedded in polypropylene with six capacitive grids
orthogonal to six inductive grids.}
\end{figure}

The combined 3-D model of the HWP used for HFSS simulations is
shown in Figure 5. The substrate material is polypropylene with
dielectric constant of 2.19. The inductive grids only interact
effectively with incident radiation polarized with its electric
field, ${\bf E}$, parallel to the strips ({\bf E} parallel to the
x-axis) and the capacitive grids only interact effectively with
incident radiation polarized with its electric field, ${\bf E}$,
parallel to the capacitive strips (${\bf E}$ parallel to the
y-axis). The radiation with {\bf E} parallel to the y axis is
transmitted through the inductive grids with a phase delay due to
the optical path length though the dielectric alone. Similarly,
the radiation with ${\bf E}$ parallel to the x-axis is transmitted
through the capacitive grids with a phase delay due to the optical
path length through the dielectric alone. The full design
parameters for HFSS models are shown in Table 1.

\begin{table}[h!]
{\bf \caption{Design parameters in HFSS: The geometric dimensions
of all spacers and the mesh patterns}}

\begin{center}
\begin{tabular}{lccccc}\hline%{|l{3.5in}|c{1.5in}|c{1.0in}|c{1.0in}|}\hline
    {\bf Parameter} & Value & Units \\
    \hline
    Polypropylene thickness above the capacitive layers(Top)     & $286$     & $\mu$m \\
    \hline
    Space between the capacitive layers                          & $155.9$ & $\mu$m \\
    \hline
    Space between the inductive layers                           & $329$   & $\mu$m \\
    \hline
    Polypropylene thickness below the inductive layers(bottom)   & $154$   & $\mu$m \\
    \hline
    Inductive grids pattern:distance between the lines           & $750$   & $\mu$m \\
    \hline
    Inductive grids pattern:line width                           & $26$    & $\mu$m \\
    \hline
    Capacitive grids pattern:distance between the lines          & $100$   & $\mu$m \\
    \hline
    Capacitive grids pattern:line width                          & $10$    & $\mu$m \\
    \hline
    Inter Digital Capacitor(IDC) pattern ($^a$)                     &     &  \\
    \hline
    Capacitive grids pattern:distance between two IDCs            & $150$   & $\mu$m \\
    \hline
\end{tabular}
\end{center}
($^a$) \small{The detailed pattern is shown in Figure 2(c).}
\end{table}

\begin{figure}[h!]\label{Fig-6}
\centering
\includegraphics[width=12cm,height=6.5cm,angle=0]{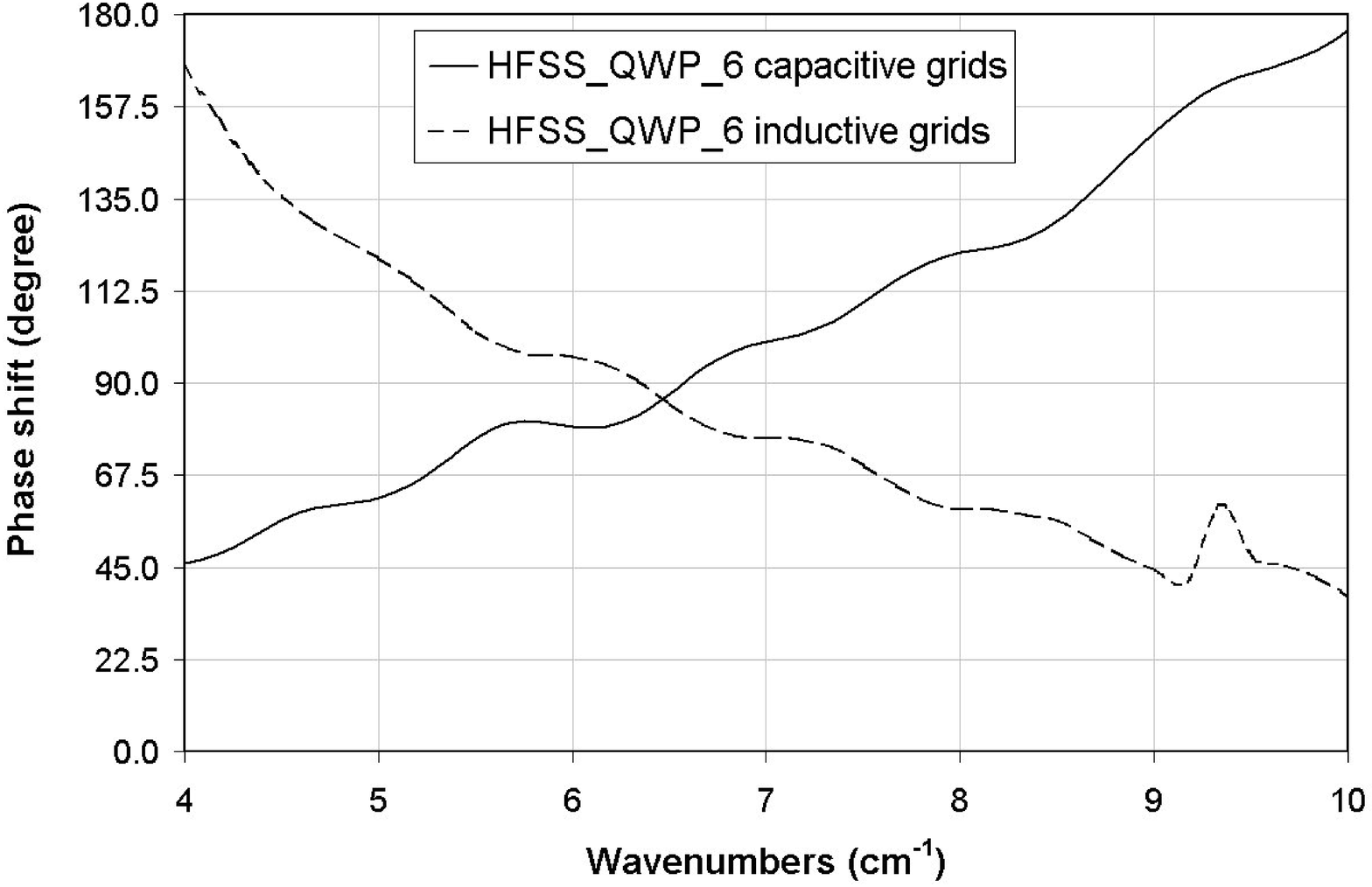}
\caption{HFSS simulated phase shift of 6 capacitive grids (first
quarter-wave plate [QWP]) and 6 inductive grids (second QWP).}
\end{figure}

The phase shift is shown in Figure 6 as a function of frequency of
the transmitted radiation, compared to the incident radiation
through either the combined six capacitive grids embedded in
dielectric or the combined six inductive grids embedded in
dielectric, calculated from the HFSS model. In both cases, the
common phase shift due to the path length through the dielectric
substrate has been subtracted. The set of capacitive grids and the
set of inductive grids were modelled separately due to the
computer memory limit. The overall phase shift of the HWP is the
sum of the two phase shifts and is very close to 180 degrees. The
phase is frequency dependent so for transmission along the
non-reactive axes we expect a linear phase dependence as shown in
Figure 7 with respect to the incidence wave. The frequency
dependence of the phases for the reactive axes are also shown in
Figure 7. The total simulated transmission and phase shift from
both ADS and HFSS are compared in Figure 3 and 4. It can be seen
that the results from both packages are in very good agreement.

\begin{figure}[h!]\label{Fig-7}
\centering
\includegraphics[width=12cm,height=9cm,angle=0]{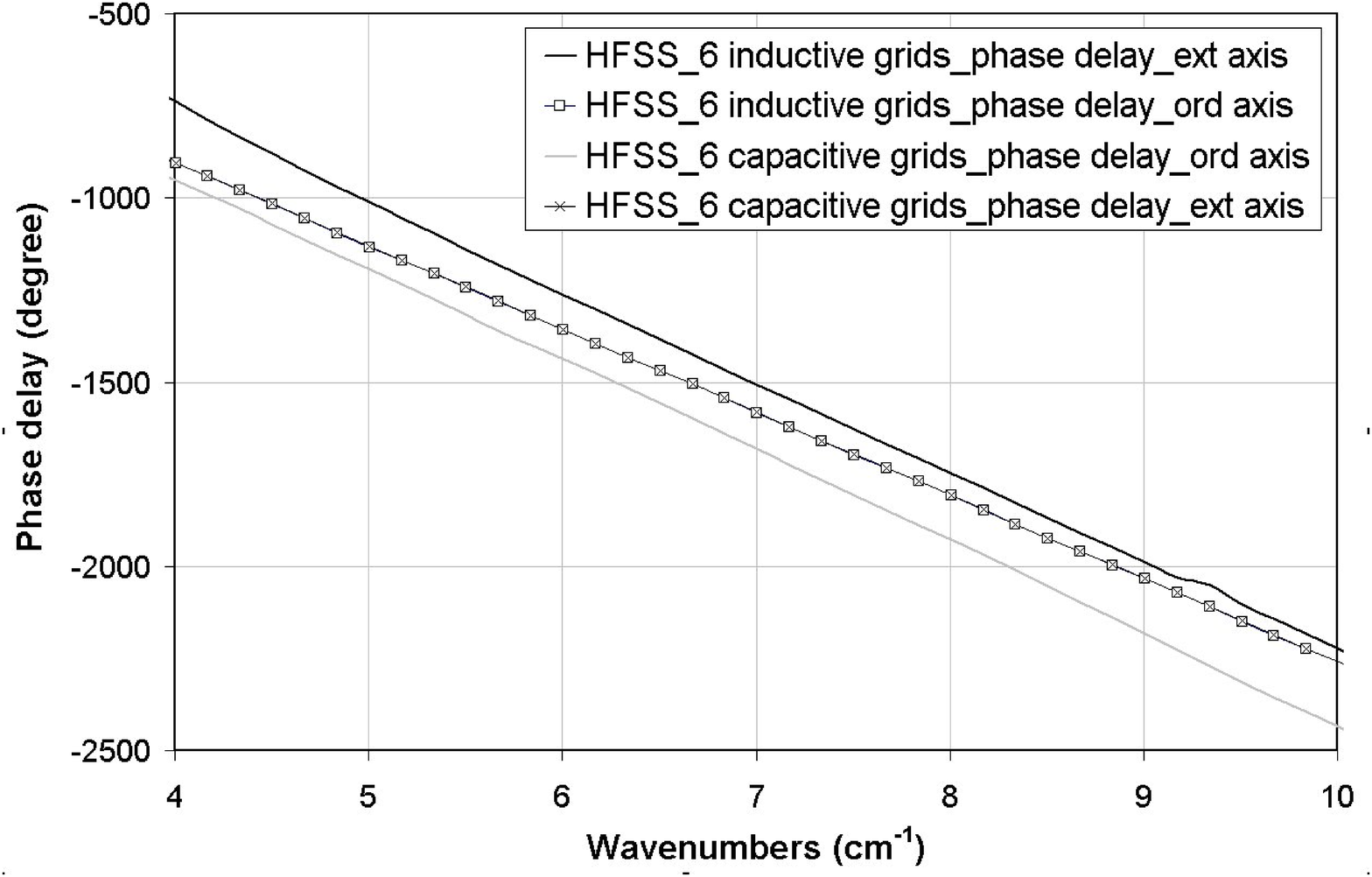}
\caption{HFSS simulated phase delay of 6 capacitive grids (first
QWP) and 6 inductive grids (second QWP) in both ordinary axis and
extraordinary axis.}
\end{figure}

\section{HWP manufacture}

Fabrication of the prototype HWP plate follows steps used in the
manufacturing of far-infrared low-pass filters \cite{Tucker2007}.
In summary, a photolithographic process is used to produce the
metal mesh patterns in copper deposited onto thin substrates
(8\,$\mu$m polypropylene). The multi-layer structure is then
assembled using additional non-metallized dielectric layers to
create the appropriate spacing between the elements. It is
particularly important to maintain good rotational and
translational alignment between the layers as this will affect the
final performance. Having assembled all of the meshes, with the
inductive and capacitive layers orientated orthogonally to each
other, the whole ensemble is then fused by the hot-pressing
process used in standard filter production \cite{Tucker2007}.

\section{Measurements and data analysis}
\subsection{Measurements of the prototype HWP}

To measure the performance of the final HWP we used a polarizing
Fourier Transform Spectrometer (pFTS hereafter). The schematic
drawing of the measurement configuration is shown in Figure 8. The
pFTS has a natural output focused beam $f_{\#}=3.5$, or more
simply a converging beam with angles $\theta \le 8^{\circ}$. To
avoid averaging the transmission and phase shift response of the
HWP over this range of input angles, a quasi-parallel beam section
was created with the use of two planar convex polyethylene lenses.
The maximum range of incident angles is then limited by the input
source aperture of the pFTS mercury arc lamp source (10\,mm), a
beam spread of only 1.6$^{\circ}$. Further, the HWP is placed
centrally in the collimated beam section between two polarizers
placed at a distance such that they can be easily tilted with
respect to the optical axis, as shown in Figure 8, to avoid
standing wave effects. The efficiency of these polarizers was
separately determined to exceed 99.8\,\% over the range of
frequencies of interest (50--1000\,GHz) with a cross polarization
of less than 0.1\,\%. The polarizers are initially aligned with
respect to each other with the grid wires horizontal and are also
aligned with respect to the optical bench in order to avoid any
projection effect when tilted.

\begin{figure}[h!]\label{Fig-8}
\centering
\includegraphics[width=12cm,height=8cm,angle=0]{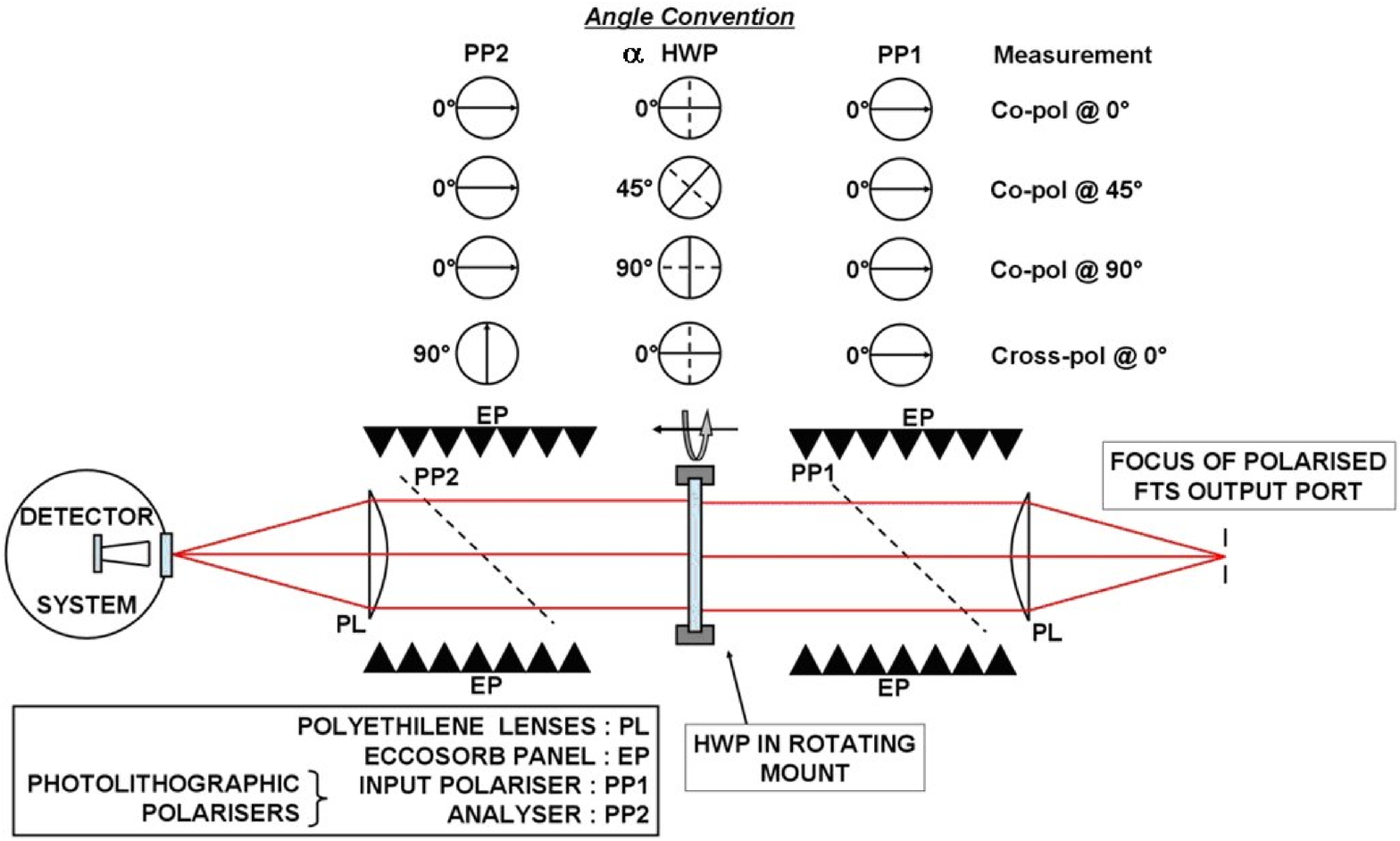}
\caption{The schematic drawing of spectral measurements set up.}
\end{figure}

A first preliminary data set is obtained by scanning the
spectrometer in the absence of the HWP, which we will refer to as
the background spectrum. This data set is fundamental as it is the
set against which all following data sets are divided in order to
remove the spectral characteristics of the source, FTS optics and
detector system. Subsequently the HWP is inserted in between the
tilted polarizers and data is acquired. Each ``set'' of data
consists of an average of 6 spectra each obtained by computing the
Fourier transform of an average of 10 interferograms with the
mirror scanned in both the forward and backward directions. The
processing pipeline uses a non-linear phase correction algorithm
\cite{Forman1966} and removes any residual asymmetry in the data
taken when the mirror is scanned in the forward and backward
directions by averaging the two data sets to cancel effects from
the finite detector time response.

The bolometric detector was cryogenically filtered to provide
spectral coverage from 60 to 660\,GHz to minimize photon noise.
All data were recorded with a spectral resolution of
$\sim$1.5\,GHz set by the pFTS optical path difference. The
transmission spectra are obtained by ratioing the sample spectra
against the background spectra to determine the HWP response
alone. The HWP rotation was controlled by using a digital angular
encoder on a large aperture annular rotator which held the plate.

An initial task was to verify that the setup is symmetric with
respect to the HWP rotation. To do this, measurements were taken
along one axis of the HWP and then repeated after rotation by 180
degrees. Theoretically, due to polarization symmetry, no change
should be observed in the pairs of data, as shown in Figure 9 for
the two HWP axes. The fact that the curves are superimposed
verifies that there are no artifacts arising from a misalignment
of the axis of rotation of the HWP rotator with the optical axis
of the spectrometer.

\begin{figure}[h!]\label{Fig-9}
\centering
\includegraphics[width=12cm,height=8cm,angle=0]{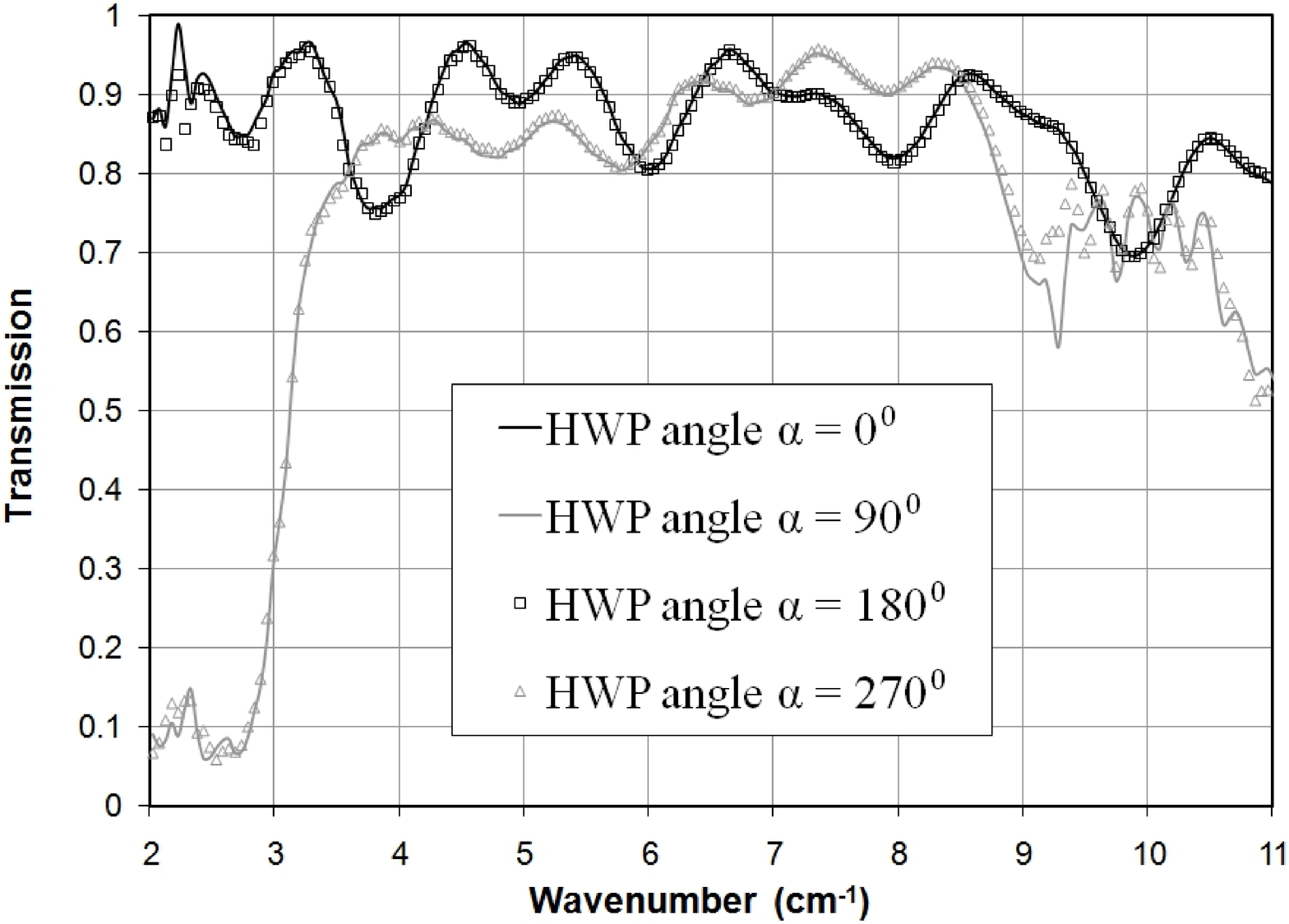}
\caption{The measured transmission between parallel polarizers
through the prototype HWP.}
\end{figure}

Having identified the polarization axes positions with respect to
the mechanical rotator, the HWP was rotated to the null positions
of the co-polar setup (Figure 8) and measured in both the co-polar
and cross-polar response at +45 and -45 degrees again verifying
that there were no measurable systematic effects. Unlike the
co-polarization maxima where symmetry is effectively broken by the
incoming polarization either being parallel to the inductive or
the capacitive grids, all the null degree positions are expected
to perform in an identical manner. This is indeed the case as can
be seen from the data in Figure 10. The curves are again
superimposed in pairs to show that the symmetry of the phase shift
is preserved and that the two sets of capacitive and inductive
grids are perpendicularly aligned with good precision.

\begin{figure}[h!]\label{Fig-10}
\centering
\includegraphics[width=12cm,height=8cm,angle=0]{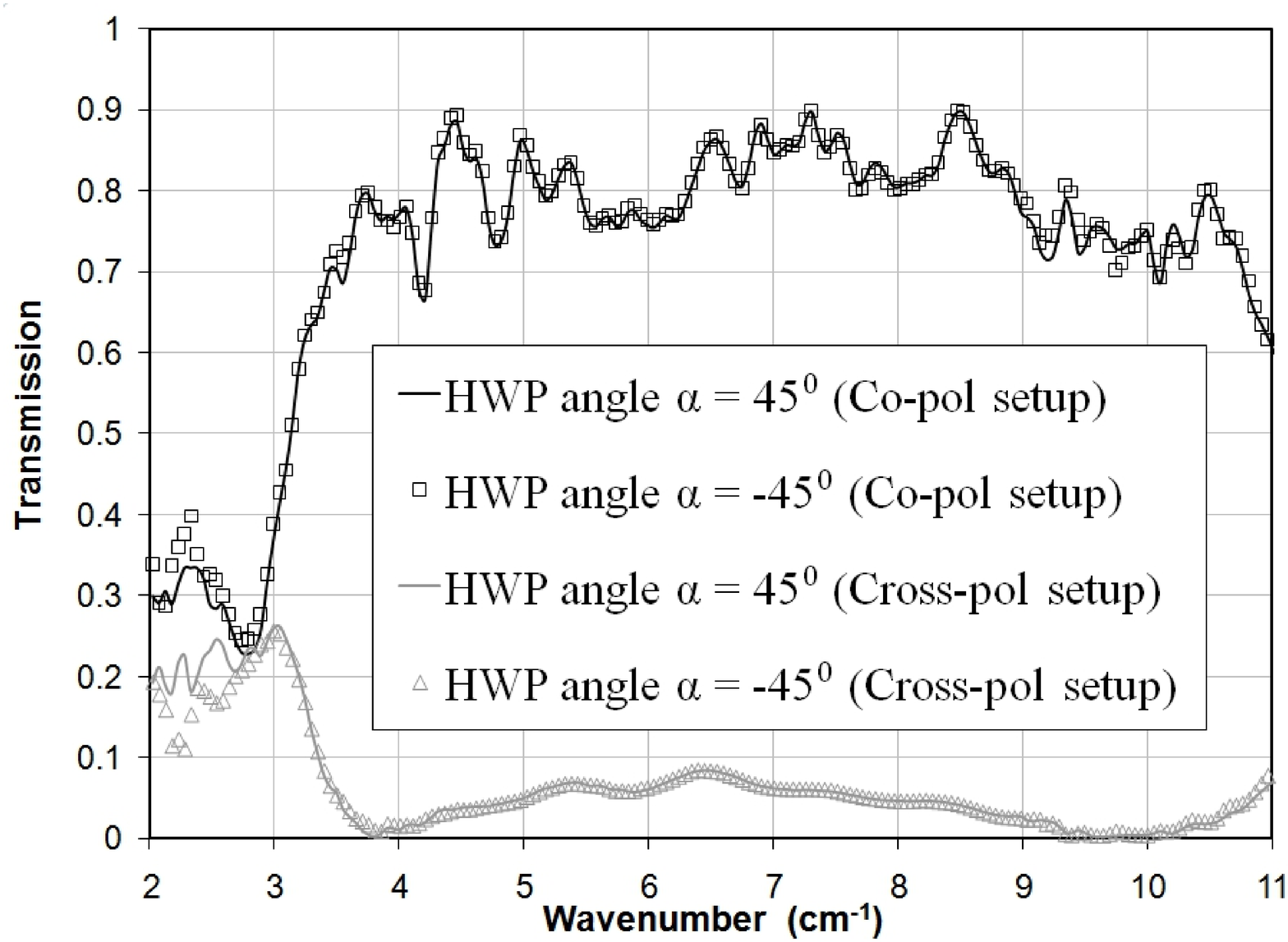}
\caption{The transmission of the co-polarization and
cross-polarization of an incoming field at 45 degrees and -45
degrees.}
\end{figure}

%The transmission of the HWP at $\alpha = 45^{\circ}$ (co-polar) is expected to be lower when compared to that of previously built crystal waveplates. Furthermore,
The cross-pol transmission on axis (0 and 90 degrees) is shown in
Figure 11. Similarly to a single crystal plate cut along its
principal axis, there should be no polarization rotation. The fact
that the transmission here is less than 0.1\,\% (corresponding to
the noise level) points to a good alignment of all the grids in
the prototype.

\begin{figure}[h!]\label{Fig-11}
\centering
\includegraphics[width=12cm,height=8cm,angle=0]{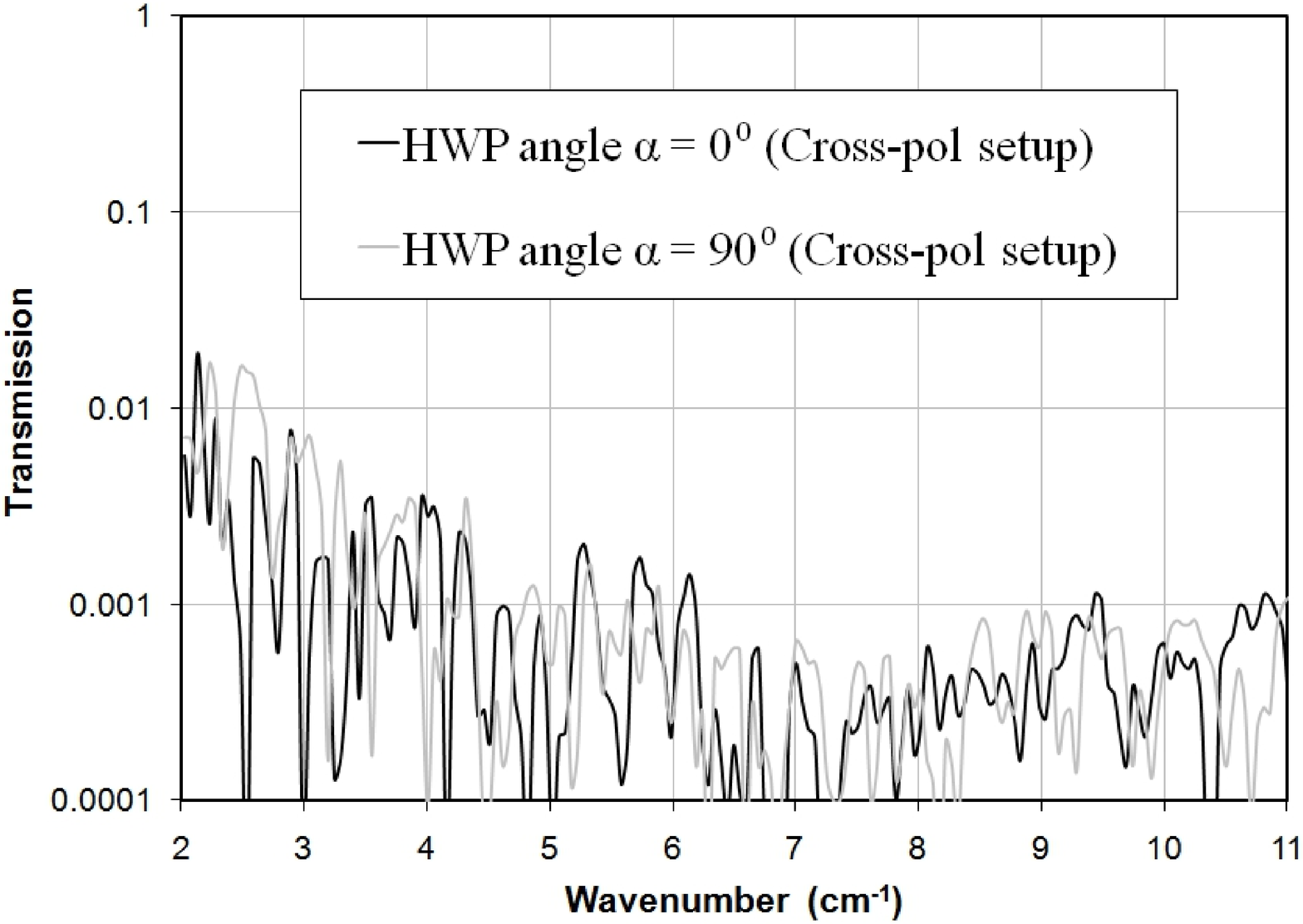}
\caption{The measured cross-polarization transmission on axis
through prototype HWP.}
\end{figure}

%Finally, to appreciate how effective the HWP is in modulating the polarized input beam from the pFTS, data were recorded as a function of the HWP rotational orientation. For graphical clarity, only subsets of these data are shown in Figure xxx.

\subsection{Discussion of HWP performance}

The average transmissions for linear polarization along the two
axes of the HWP assembly are between 86 and 91 percent, the
precise value being determined by the spectral range that is
considered within the [105,315]\,GHz interval where the phase
shift is the most effective. These transmissions, shown in Figure
12, match closely the HFSS simulations up to ($\sim$300\,GHz)
where diffraction starts.

\begin{figure}[h!]\label{Fig-12}
\centering
\includegraphics[width=12cm,height=8cm,angle=0]{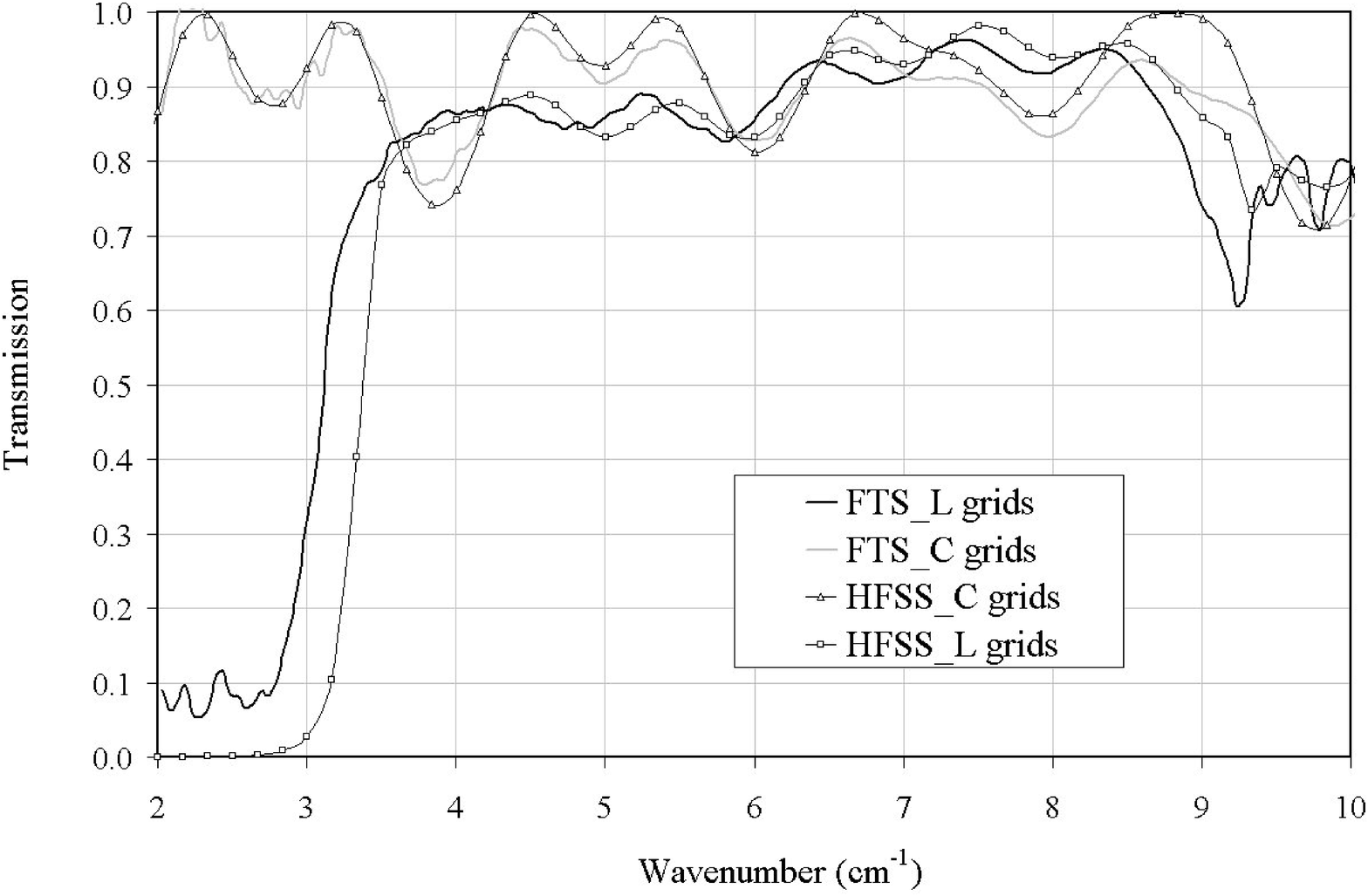}
\caption{Measured (FTS) and HFSS simulated power transmission
through prototype HWP when the electric field is parallel to the
inductive lines (inductive grids) and orthogonal to it (capacitive
grids).}
\end{figure}

The simulation includes measurements of the absorption loss in the
polypropylene substrate material \cite{Tucker2007} and a
resistivity for copper which is increased to represent an
evaporated rather than bulk material. The fit to the data is seen
to be good for the capacitive axes as shown by the overlay in
Figure 12, but not so good for the inductive axis. We believe that
the poor fit here has several origins. First the inductive strips
could be slightly over-etched in the manufacture process which
would lead to the low frequency edge moving to shorter frequencies
as observed. Secondly, the copper thickness of $\sim$0.1\,$\mu$m
is equivalent to the skin depth near 1\,cm$^{-1}$ which could
explain the non-zero rejection at low frequencies and associated
phase errors. Finally, there could be systematic effects arising
from the non-perfect alignment of the inductive and capacitive
elements during manufacture, producing an additional source of
reflection (due to impedance mismatch). Currently HFSS cannot cope
with large open structures for which the symmetry is broken.
However, the fact that this prototype HWP is seen to perform well
shows that these systematic effects are small.

One of the key points of polarization modulation is the leakage
between axes. To be effective, the phase delay between the two HWP
axes should be as close to $\pi$ as possible to avoid transforming
linear polarization in elliptical, hence losing efficiency when
used as a linear polarization modulators. The phase cannot be
directly measured in a pFTS but it can be indirectly inferred from
the expected value of the Mueller Matrix of the optical element.

The Mueller matrix of a phase-delaying device with non-identical
axes transmission can be written in its most simple form, when the
reference frame is aligned with the plate axes, as:

\begin{equation}
M_{\rm HWP}(0^{\circ})=\frac{1}{2}
  \begin{pmatrix}
    \alpha^2 + \beta ^2 & (\alpha^2 - \beta ^2) & 0 & 0 \\

    (\alpha^2 - \beta ^2) & (\alpha^2 + \beta ^2) & 0 & 0 \\
    0 & 0 & 2\alpha\beta\cos\phi & 2\alpha\beta\sin\phi \\
    0 & 0 & -2\alpha\beta\sin\phi & 2\alpha\beta\cos\phi
  \end{pmatrix}
\end{equation}

We have used the notation presented in \cite{Savini2009} where
$\alpha^2$ and $\beta^2$ represent the intensity transmissions on
the plate's two axes and $\phi$ is the phase shift between the two
axes. Here we are interested in extracting the frequency
dependence of these three quantities from the spectral
measurements.

Recovering the first two parameters is intuitively straightforward
by aligning the input and output polarizers and measuring the
plates' spectral transmission. Extracting the phase $\phi$
requires a combination of measurements. First consider the Mueller
matrix of the HWP rotated by 45 degrees:

\begin{equation}
M_{\rm HWP}(45^{\circ})=\frac{1}{2}
  \begin{pmatrix}
    \alpha^2 + \beta ^2 & 0 & (\alpha^2 - \beta ^2) & 0 \\
    0 & 2\alpha\beta\cos\phi & 0 & -2\alpha\beta\sin\phi \\
    \alpha^2 - \beta ^2 & 0 & \alpha^2 + \beta ^2 & 0 \\
    0 & 2\alpha\beta\sin\phi & 0 & 2\alpha\beta\cos\phi
  \end{pmatrix}
\end{equation}

\begin{figure}[h!]\label{Fig-13}
\centering
\includegraphics[width=12cm,height=8cm,angle=0]{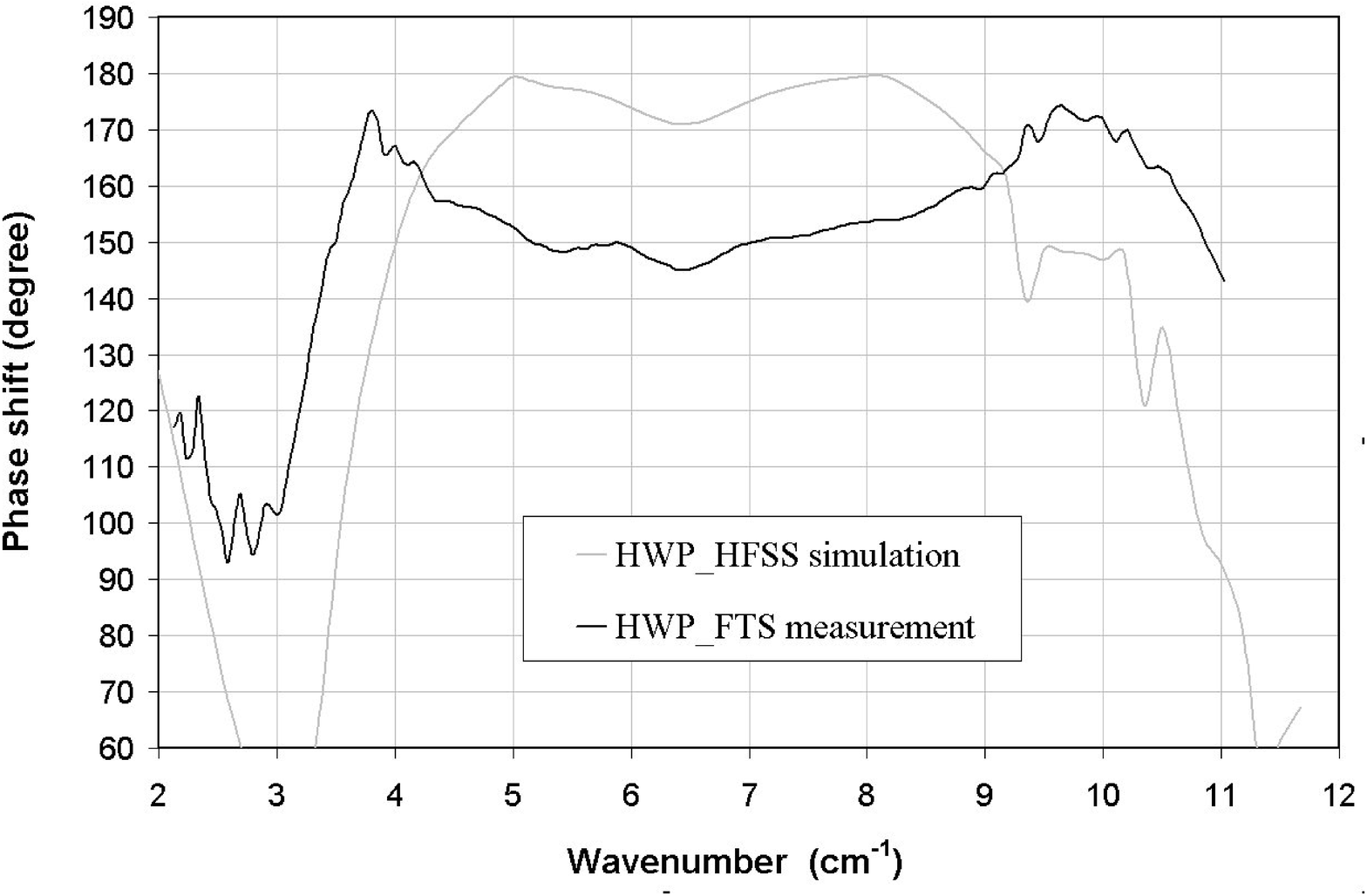}
\caption{The HFSS simulated and measured phase shift.}
\end{figure}

This allows measurements performed with Stokes vectors Q to
extract a dependence on the phase. By noting a set of data taken
with the HWP at an angle $\alpha$ between parallel polarizers
(Co-pol in Figure. 8) as $D(\alpha)$ we can combine the sets of
measurements in the following way to obtain the phase shift as
shown in Figure 13:

\begin{eqnarray}
D(45^{\circ})=\frac{1}{4}(\alpha^2 + \beta ^2 + 2\alpha\beta\cos\phi) \\
\phi=\cos^{-1}\{[4D(45^{\circ})-D(0^{\circ})-D(90^{\circ})]/[2\sqrt{D(0^{\circ})D(90^{\circ})}]\}
\end{eqnarray}

The resulting phase shift presents a similar functional dependence
but is also substantially different quantitatively from the HFSS
simulated value. The discrepancy between expected and measured
phase shift is likely to be sought in the physical parameters that
cannot be easily introduced in the HFSS simulation due to
resulting excessively large mesh geometries. One of these
parameters is the thickness of the copper deposition on the metal
mesh layers. A copper thickness of one hundred nanometers is of
the same order of the skin depth of the metal at the frequencies
of interest. As the skin depth scales as $1/\sqrt{\omega}$, the
additional phase delay produced by delayed currents induced on the
other side of the metal meshes will vary with frequency across the
band. A crude estimation of the absorption can be obtained by
considering the envelope of the parallel polarizer transmission
curves as a smooth and monotonic function of frequency. This shows
that an overall absorption of 3 to 5 percent can be expected from
the combined absorption of the polypropylene substrate and the
loss due to the imperfect conductivity of the metal.

\section{Conclusions}

It is important to put the performance of this HWP into context
with the current crystalline HWPs. The two represent very
different technologies and both have manufacture and performance
issues. Here we attempt to summarize these differences to aid in
their selection for future instruments.

Although the prototype has been designed and built at
mm-wavelengths, its broadband nature and impedance characteristics
are conserved when scaling the geometries to operate at higher
frequencies (at sub-mm wavelengths and frequencies higher than
1\,THz), where crystalline absorption is indeed a problem. A
return loss of about ten percent is experienced for this prototype
metal mesh HWP without applying additional anti-reflection
coating; its average transmission is an improvement with respect
to most crystalline birefringent waveplates which have a large
impedance mismatch due to the high index of refraction of the
crystal material. Single layer anti-reflection coatings, while
improving the insertion loss of the crystal alternatives, cannot
avoid the inherent crystalline absorption, which gets
progressively worse at higher frequencies \cite{Lau2006}. While
reflections can present a problem due to ghost images appearing in
a polarimeter system which has more than one reflecting element,
absorption is a more serious problem for low background sub-mm
astronomical instruments, since the detectors employed in these
are background photon noise limited. Thermal grey body emission
from a room temperature HWP, which for low loss is proportional to
its absorption, can present a significant additional noise
component. Most experiments are thus forced to position the HWP
modulator inside the cryogenic housing at temperatures below 100K,
since the absorption decreases and the grey body emission is also
reduced.

While cryogenic deployment of the hot-pressed sample in question,
which is unnecessary, would not constitute a problem, in the case
of AR coated crystals large temperature gradients require great
attention and planning in both the construction process and in the
mount design.

Although the recovered phase is not an improvement on existing
achromatic crystalline devices \cite{Savini2006, Savini2009}, the
modulation efficiency of the HWP is always above 85\,\% in a
90\,\% spectral bandwidth with the best performance above 96\,\%
modulation efficiency in two 10\,\% narrow bands (centered at 125
and 295\,GHz) at the extremes of its spectral range.

In conclusion we note that overall there are numerous instrumental
advantages of this device that make it a competitive alternative
to crystalline HWP, especially at higher frequencies and where
wide spectral bandwidth is required.

\bibliographystyle{unsrt}
\end{document}